\documentclass[aps,prd,twocolumn,
comment, 
superscriptaddress,groupedaddress,
amsmath,amssymb,floatfix,nofootinbib]
{revtex4}  
\usepackage{float}
\usepackage{graphicx}
\usepackage{bm}
\usepackage[usenames,dvipsnames]{color}
\usepackage{slashed}
\usepackage{slashed}

\begin{document}

%

\let\a=\alpha      \let\b=\beta       \let\c=\chi        \let\d=\delta
\let\e=\varepsilon \let\f=\varphi     \let\g=\gamma      \let\h=\eta
\let\k=\kappa      \let\l=\lambda     \let\m=\mu
\let\o=\omega      \let\r=\varrho     \let\s=\sigma
\let\t=\tau        \let\th=\vartheta  \let\y=\upsilon    \let\x=\xi
\let\z=\zeta       \let\io=\iota      \let\vp=\varpi     \let\ro=\rho
\let\ph=\phi       \let\ep=\epsilon   \let\te=\theta
\let\n=\nu
\let\D=\Delta   \let\F=\Phi    \let\G=\Gamma  \let\L=\Lambda
\let\O=\Omega   \let\P=\Pi     \let\Ps=\Psi   \let\Si=\Sigma
\let\Th=\Theta  \let\X=\Xi     \let\Y=\Upsilon
%


\def\cA{{\cal A}}                \def\cB{{\cal B}}
\def\cC{{\cal C}}                \def\cD{{\cal D}}
\def\cE{{\cal E}}                \def\cF{{\cal F}}
\def\cG{{\cal G}}                \def\cH{{\cal H}}
\def\cI{{\cal I}}                \def\cJ{{\cal J}}
\def\cK{{\cal K}}                \def\cL{{\cal L}}
\def\cM{{\cal M}}                \def\cN{{\cal N}}
\def\cO{{\cal O}}                \def\cP{{\cal P}}
\def\cQ{{\cal Q}}                \def\cR{{\cal R}}
\def\cS{{\cal S}}                \def\cT{{\cal T}}
\def\cU{{\cal U}}                \def\cV{{\cal V}}
\def\cW{{\cal W}}                \def\cX{{\cal X}}
\def\cY{{\cal Y}}                \def\cZ{{\cal Z}}


\def\be{\begin{equation}}
\def\ee{\end{equation}}
\def\bea{\begin{eqnarray}}
\def\eea{\end{eqnarray}}
\def\bm{\begin{matrix}}
\def\em{\end{matrix}}
\def\bpm{\begin{pmatrix}}
    \def\epm{\end{pmatrix}}

{\newcommand{\lsim}{\mbox{\raisebox{-.6ex}{~$\stackrel{<}{\sim}$~}}}
{\newcommand{\gsim}{\mbox{\raisebox{-.6ex}{~$\stackrel{>}{\sim}$~}}}
\def\mpl{M_{\rm {Pl}}}
\def\gev{{\rm \,Ge\kern-0.125em V}}
\def\tev{{\rm \,Te\kern-0.125em V}}
\def\mev{{\rm \,Me\kern-0.125em V}}
\def\ev{\,{\rm eV}}

\title{\boldmath  Explaining the diphoton excess in Alternative Left-Right Symmetric Model}
\author{Chandan Hati}
\email{chandan@prl.res.in} 
\affiliation{Physical Research Laboratory, Navrangpura, Ahmedabad 380 009, India}
\affiliation{Indian Institute of Technology Gandhinagar, Chandkheda, Ahmedabad 382 424, India}

\begin{abstract}
We propose a possible explanation of the recent diphoton excess reported by ATLAS and CMS collaborations, at around 750 GeV diphoton invariant mass, within the framework of $E_{6}$ motivated Alternative Left-Right Symmetric Model (ALRSM), which is capable of addressing the $B$ decay anomalies in the flavor sector, the $eejj$ and $e\slashed{p}_{T}jj$ excesses reported by CMS in run 1 of LHC and has the feature of high scale leptogenesis. We find that gluon-gluon fusion can give the observed production rate of the $750 \gev$ resonance, $\tilde{n}$, through a loop of scalar leptoquarks ($\tilde{h}^{(c)}$) with mass below a few TeV range, while $\tilde{n}$ can subsequently decay into $\gamma\gamma$ final state via loops of $\tilde{h}^{(c)}$ and $\tilde{E}^{(c)}$. Interestingly, the $\tilde{E}^{(c)}$ loop can enhance the diphoton branching ratio significantly to successfully explain the observed cross section of the diphoton signal.
\end{abstract}
\maketitle

\section{Introduction}
The CMS and ATLAS collaborations have recently announced the search results based on the first 3 ${\rm{fb}}^{-1}$ of collected data from Run 2 of the LHC at $\sqrt{s}=13$ \tev \cite{lhcrun2a,atlasconf,CMS:2015dxe}. The ATLAS collaboration has reported a 3.9 $\sigma$ local (2.3 $\sigma$ global) excess in the diphoton channel at the diphoton invariant mass of around $750 \gev$ with $3.2 \rm{fb}^{-1}$ integrated luminosity. This excess corresponds to about 14 events appearing in at least two energy bins, suggesting a large width $\sim 45 \gev$, however more data is required to confirm the existence of this feature. The CMS collaboration has partially endorsed this result with an integrated luminosity of  $2.6 \rm{fb}^{-1}$. They have reported about $10$ excess events in the $\gamma \gamma$ channel peaked at $760 \gev$ amounting to a $2.6 \sigma$ local ($<1.2 \sigma$ global) excess. 

A new resonance coupling with the Standard Model (SM) $t$ quark or $W^{\pm}$ can give rise to loop diagrams with $\gamma \gamma$ final state, however, such diagrams are highly suppressed at the large $\gamma \gamma$ invariant masses and the dominant decay channels are $t \bar{t}$ or $W^{+}W^{-}$. Thus, the observation of the $\gamma \gamma$ resonance at 750 $\gev$ (much larger than the electroweak symmetry breaking scale) presumably hints towards new physics around that mass scale. Several new physics interpretations of the diphoton signal have been proposed in the literature explaining the excess events 
\cite{DiChiara:2015vdm,Pilaftsis:2015ycr,Knapen:2015dap,Backovic:2015fnp,Molinaro:2015cwg,Gupta:2015zzs,
Ellis:2015oso,Higaki:2015jag,Mambrini:2015wyu,Buttazzo:2015txu,Franceschini:2015kwy,Angelescu:2015uiz,
Bellazzini:2015nxw,McDermott:2015sck,Low:2015qep,Petersson:2015mkr,
Cao:2015pto,Kobakhidze:2015ldh,Agrawal:2015dbf,Chao:2015ttq,Fichet:2015vvy,Curtin:2015jcv,Csaki:2015vek,
Aloni:2015mxa,Demidov:2015zqn,No:2015bsn,Bai:2015nbs,Matsuzaki:2015che,Dutta:2015wqh,Becirevic:2015fmu,
Cox:2015ckc,Martinez:2015kmn,Bian:2015kjt,Chakrabortty:2015hff,Ahmed:2015uqt,
Deppisch:2016scs,Hernandez:2016rbi,Dutta:2016jqn,Modak:2016ung,
Danielsson:2016nyy,Chao:2016mtn,Csaki:2016raa,Karozas:2016hcp,
Ghorbani:2016jdq,Han:2016bus,Ko:2016lai,Nomura:2016fzs,
Ma:2015xmf,Palti:2016kew,Potter:2016psi,Jung:2015etr,
Marzola:2015xbh,Falkowski:2015swt,
Nakai:2015ptz,Harigaya:2015ezk,
Ellwanger:2015uaz,Karozas:00640,Csaki:00638,Chao:00633,Danielsson:00624,Ghorbani:00602,
Ko:00586,Han:00534,Nomura:00386,Palti:00285,Potter:00240,Palle:00618,Dasgupta:2015pbr,Bizot:2015qqo,
Goertz:2015nkp,Kim:2015xyn,Craig:2015lra,Cheung:2015cug,Allanach:2015ixl,Altmannshofer:2015xfo,Huang:2015rkj,
Belyaev:2015hgo,Liao:2015tow,Chang:2015sdy,Luo:2015yio,
Kaneta:2015qpf,Hernandez:2015hrt,Low:2015qho,Dong:2015dxw,Kanemura:2015vcb,Kanemura:2015bli,Kang:2015roj,Chiang:2015tqz,
Ibanez:2015uok,Huang:2015svl,Hamada:2015skp,Anchordoqui:2015jxc,Bi:2015lcf,Chao:2015nac,Cai:2015hzc,Cao:2015apa,
Tang:2015eko,Dev:2015vjd,Gao:2015igz,Cao:2015scs,Wang:2015omi,An:2015cgp,Son:2015vfl,Li:2015jwd,Salvio:2015jgu,
Park:2015ysf,Han:2015yjk,Hall:2015xds,Casas:2015blx,Zhang:2015uuo,Liu:2015yec,Das:2015enc,Davoudiasl:2015cuo,Cvetic:2015vit,
Chakraborty:2015gyj,Badziak:2015zez,Patel:2015ulo,Moretti:2015pbj,Gu:2015lxj,Cao:2015xjz,Pelaggi:2015knk,Dey:2015bur,
Hernandez:2015ywg,Murphy:2015kag,deBlas:2015hlv,Dev:2015isx,Boucenna:2015pav,Kulkarni:2015gzu,Chala:2015cev,
Bauer:2015boy,Cline:2015msi,Berthier:2015vbb,Kim:2015ksf,Bi:2015uqd,Heckman:2015kqk,Huang:2015evq,Cao:2015twy,Wang:2015kuj,
Antipin:2015kgh,Han:2015qqj,Ding:2015rxx,Chakraborty:2015jvs,Barducci:2015gtd,Cho:2015nxy,Feng:2015wil,Bardhan:2015hcr,
Han:2015dlp,Dhuria:2015ufo,Chang:2015bzc,Han:2015cty,Arun:2015ubr,Chao:2015nsm,Bernon:2015abk,Carpenter:2015ucu,Megias:2015ory,
Alves:2015jgx,Gabrielli:2015dhk,Kim:2015ron,Benbrik:2015fyz,Jiang:2015oms,Ito:2016zkz,Zhang:2016pip,Ma:2016qvn,Bhattacharya:2016lyg,Fichet:2016pvq,Borah:2016uoi}.
In light of the fact that the two collaborations have suggested signal events consistent with each other at $3\sigma$ statistical significance level, hinting towards a new physics scenario, it is important to explore the possible model framework that can naturally accommodate the diphoton signal. From a theoretical point of view, a framework explaining the diphoton excess along with other signals for new physics from the flavor sector and excesses reported in first run of LHC are particularly interesting. 
 
 The measurements of the branching fractions $\bar{B}\rightarrow D^{(\ast)}\tau \bar{\nu}$, $\cR(D^{(\ast)})=Br(\bar{B}\rightarrow D^{(\ast)}\tau \bar{\nu})/Br(\bar{B}\rightarrow D^{(\ast)} l \bar{\nu})$, reported by the BaBar \cite{Lees:2012xj, Lees:2013uzd} and Belle \cite{Huschle:2015rga} collaborations  and the ratio of branching fractions $\bar{B}\rightarrow D^{\ast}\tau \bar{\nu}$ reported recently by LHCb \cite{Aaij:2015yra} are consistent with each other and hint towards a new physics scenario. The BaBar and Belle collaborations have reported $\cR(D)^{{\rm BaBar}}=0.440\pm 0.058  \pm 0.042 $, $\cR(D)^{{\rm Belle}}=0.375\pm 0.064  \pm 0.026 $ and $\cR(D^{\ast})^{{\rm BaBar}}=0.332\pm 0.024 \pm 0.018 $, $\cR(D^{\ast})^{{\rm Belle}}=0.293\pm 0.038  \pm 0.015 $, with the SM expectations $\cR(D)^{{\rm SM}}=0.300\pm 0.010$ and $\cR(D^{\ast})^{{\rm SM}}=0.252\pm 0.005$. While the LHCb collaboration has reported $\cR(D^{\ast})=0.336\pm 0.027 ({\rm stat.}) \pm 0.030 ({\rm syst.})$ with the SM expectation $0.252\pm0.005$, showing a $2.1\sigma$ excess \cite{Aaij:2015yra}. These results corroborate the earlier measurements \cite{Aubert:2007dsa, Bozek:2010xy} and combined together show significant enhancements over the SM expectations. On the other hand, the LHCb has reported  a $2.6 \sigma$ deviation from the SM expectation of the ratio of the branching fractions, $\cR_{K}=Br(B^{+}\rightarrow K^{+}\mu^{+}\mu^{-})/Br(B^{+}\rightarrow K^{+} e^{+} e^{-})$, in the invariant mass region, $1 \gev^{2}\leq M_{ll} \leq 6\gev^{2}$, of the dilepton pair with an integrated luminosity of $3 \rm{fb}^{-1}$ \cite{Aaij:2014ora}. Several scenarios involving color triplet scalar leptoquarks have been proposed explaining the above anomalies and other issues such as anomalous muon magnetic moment \cite{Fajfer:2012jt,Tanaka:2012nw,Dorsner:2013tla,Sakaki:2013bfa,Hiller:2014yaa,Sahoo:2015wya,Freytsis:2015qca,Bauer:2015knc,Sahoo:2015pzk}. Thus a model framework which naturally accommodates scalar leptoquarks are particularly attractive from a flavor physics stand point.
 
The CMS results of search for di-leptoquark production at $\sqrt{s}=8 \rm{TeV}$ and $19.6 \rm{fb}^{-1}$ of integrated luminosity have been reported to show a $2.4\sigma$ in the $eejj$ channel and a $2.6\sigma$ local excess in the $e\slashed{p}_{T}jj$ channel, also hint towards the existence of scalar leptoquarks \cite{CMS:2014qpa}. On the other hand, in the first run of LHC, the CMS Collaboration had reported the results for the right-handed gauge boson $W_R$ search at $\sqrt{s}= 8\tev$ and $19.7 {\rm{fb}}^{-1}$ of integrated luminosity \cite{Khachatryan:2014dka}, showing a $2.8\sigma$ local excess in the $eejj$ channel at an invariant mass of $1.8 \tev < m_{eejj} < 2.2 \tev$, hinting towards a right handed $W$ boson with mass around $\sim 2 \tev$. The Left-Right Symmetric Model (LRSM) framework with $g_{L}\neq g_{R}$ embedded into UV completed higher gauge groups can explain such signal \cite{Deppisch:2014qpa,Deppisch:2014zta,Dev:2015pga,Deppisch:2015cua}, while for $g_{L}= g_{R}$ case the signal has been explained by taking into account the $CP$ phases and non-degenerate masses of heavy neutrinos in Refs. \cite{Gluza:2015goa, Jelinski:2015ifw}. Other scenarios addressing the above excesses have been proposed in Refs. \cite{Dobrescu:2014esa, Aguilar-Saavedra:2014ola, Allanach:2014lca, Queiroz:2014pra, Biswas:2014gga, Allanach:2014nna, Allanach:2015ria, Dhuria:2015hta, Dutta:2015osa, Dobrescu:2015asa, Berger:2015qra, Krauss:2015nba, Dhuria:2015swa}.

One of the attractive mechanisms to explain the baryon asymmetry of the universe is the leptogenesis mechanism  \cite{Fukugita:1986hr}. The seesaw mechanism \cite{Minkowski:1977sc,GellMann:1980vs,Yanagida:1979as,Mohapatra:1979ia,Mohapatra:1980yp} which naturally explains the smallness of neutrino masses, can generate a lepton asymmetry (and hence a $B-L$ asymmetry) via heavy neutrino (Higgs triplet) decay at a high scale, which then gets converted to the baryon asymmetry of the universe via  $B+L$ violating anomalous processes in equilibrium before the electroweak phase transition. In the standard LRSM framework, the observation of a $2 \tev$ $W_{R}$ boson at the LHC will rule out the possibilities of high scale as well as $\tev$ scale resonant leptogenesis with the standard LRSM field content due to the unavoidable fast gauge mediated $B-L$ violating interactions \cite{Ma:1998sq,Frere:2008ct,Deppisch:2015yqa,Deppisch:2013jxa,Dev:2014iva,Dhuria:2015wwa,Dhuria:2015cfa,Dev:2015vra}. Thus a framework which can explain the $2\tev$ $eejj$ signal while simultaneously accommodate a successful leptogenesis is welcome from a cosmological point of view.

In this work, we argue that the $E_{6}$ motivated Alternative Left-Right Symmetric Model (ALRSM), first proposed by Ma (1986) \cite{Ma:1986we}, provides a very attractive framework to address the diphoton excess along with the flavor anomalies reported in $B$ decays, other LHC run 1 excesses mentioned above and high scale leptogenesis. We find that gluon-gluon fusion can account for the observed production rate of the $750 \gev$ resonance, through a loop of scalar leptoquark $\tilde{h}^{(c)}$ with mass below few $\tev$ range. The scalar resonance $\tilde{n}$ can subsequently decay into $gg$ and $\gamma\gamma$ final states via the loops of $\tilde{h}^{(c)}$ and slepton $\tilde{E}^{(c)}$. Considering only scalar leptoquarks in the decay loop of $\tilde{n}$ gives a diphoton branching ratio suppressed by a factor of $10^{-3}-10^{-4}$, however, the contribution from the $\tilde{E}^{(c)}$ loop enhances the diphoton branching ratio significantly and yields the observed cross section of the diphoton signal. While this model naturally accommodates scalar leptoquarks, which can address the flavor anomalies, it can also explain the $eejj$ and $e\slashed{p}_{T}jj$ excess signals by the pair production of scalar leptoquarks or the resonant production of a slepton, and provide an attractive mechanism of high scale leptogenesis. Thus, ALRSM provides a very promising framework connecting the diphoton excess to the existent flavor and LHC run1 anomalies along with the possibility of explaining the observed baryon asymmetry of the universe.

The plan of rest of this paper is as follows. In section \ref{sec2}, we briefly introduce the Alternative Left-Right symmetric framework. In section \ref{sec2.1}, we explain diphoton signal from $\tilde{n}$ decay in ALRSM and discuss the relevant bounds on scalar leptoquark and slepton masses. In section \ref{sec3}, we review the possible explanation of the enhanced $\bar{B}\rightarrow D^{(\ast)}\tau \bar{\nu}$ decay rates, the lepton non-universality of the ratio $\cR_{K}=Br(B^{+}\rightarrow K^{+}\mu^{+}\mu^{-})/Br(B^{+}\rightarrow K^{+} e^{+} e^{-})$ in ALRSM using a scalar leptoquark. We also briefly discuss the possibility of addressing $eejj$ and $e\slashed{p}_{T}jj$ excesses reported by the CMS in the framework of ALRSM and the possible high scale leptogenesis mechanism. In section \ref{sec4}, we summarize and make concluding remarks.
\section{Alternative Left-Right Symmetric Model in a nutshell}
\label{sec2}
One of the maximal subgroups of superstring inspired $E_{6}$ group is $SU(3)_{C}\times SU(3)_{L} \times SU(3)_{R}$, under which the fundamental $27$ representation of $E_{6}$ decomposes as
\be{\label{1.1}} 
27= (3, 3, 1)+(3^{\ast}, 1, 3^{\ast})+(1, 3^{\ast},3),
\ee
where $(3, 3, 1)$ corresponds to $(u, d, h)$, $(3^{\ast}, 1, 3^{\ast})$ corresponds to $(h^{c}, d^{c}, u^{c})$ and $(1, 3^{\ast} ,3)$ corresponds to the color singlet superfields. Here $h$ represents the $-\frac{1}{3}$ charge leptoquark superfield which carries baryon number $B=\frac{1}{3}$ and lepton number $L=1$. $u, d$ corresponds to the usual up and down quark superfields. In addition to the usual leptonic superfields, the color singlet sector hosts other exotic superfields $N^{c}$, $n$ and two isodoublets $(\nu_{E}, E)$, $(E^{c},N_{E}^{c})$. These new exotic fields makes the phenomenology of the low energy subgroups of $E_{6}$ very rich and interesting. The first family of superfields corresponding to the fundamental representation are assigned as
\be{\label{1.2}}
 \bpm u \\ d \\ h \epm + \bpm u^{c} & d^{c} &
h^{c}\epm +\bpm E^{c} & \nu & \nu_{E} \\ N^{c}_{E} & e & E \\
e^{c} & N^{c} & n \epm,
\ee
where $SU(3)_{L}$ operates vertically and $SU(3)_{(R)}$ operates horizontally. Now, the $SU(3)_{(L,R)}$ further break into $SU(2)_{(L,R)} \times U(1)_{(L,R)}$ giving three choices for assigning the isospin doublets corresponding to $T, U, V$ isospins of $SU(3)_{R}$. The breaking of $SU(3)_{L}$ to $SU(2)_{L}$ is fixed by the well established SM gauge structure. When $(d^{c}, u^{c})_L$ is assigned as the $SU(2)_{R}$ doublet, the usual left-right symmetric extension of the standard model including the exotic particles results, while in another choice, corresponding to $(h^{c}, d^{c})$ assigned to the $SU(2)_{R}$ doublet \cite{London:1986dk} with the charge equation $Q=T_{3L}+\frac{1}{2} Y_{L}+\frac{1}{2} Y_{N}\,,$ gives rise to a $SU(2)_{R}$ sector which does not contribute to electric charge and is often denoted by $SU(2)_{N}$. The third possibility corresponding to the choice $(h^{c}, u^{c})$ as the $SU(2)_R$ doublet gives the subgroup referred to as the Alternative Left-Right Symmetric Model (ALRSM), first proposed by Ma (1986) \cite{Ma:1986we}. 

In ALRSM, the superfields transforms under the subgroup $G=SU(3)_{c}\times SU(2)_{L}\times SU(2)_{R^{\prime}}\times U(1)_{Y^{\prime}}$ as 
\bea {\label{1.3}}
(u, d)_{L} &:& (3, 2, 1, \frac{1}{6})\quad
(h^{c}, u^{c})_{L} : (\bar{3}, 1, 2, -\frac{1}{6})\nonumber\\
(\nu_{E}, E)_{L} &:& (1, 2, 1, -\frac{1}{2})\quad
(e^{c}, n)_{L} : (1, 1, 2, \frac{1}{2})\nonumber\\
h_{L} &:& (3, 1, 1, -\frac{1}{3})\quad
d^{c}_{L} : (\bar{3}, 1, 1, \frac{1}{3})\nonumber\\
\bpm \nu_{e} & E^{c} \\ e & N^{c}_{E}\epm_{L} &:& (1, 2, 2, 0)\quad\quad
N^{c}_{L} : (1, 1, 1, 0),
\eea
with $Y^{\prime}=Y_{L}+Y_{R}^{\prime}$. The electric charge equation is given by $Q=T_{3L}+\frac{1}{2} Y_{L}+T^{\prime}_{3R}+\frac{1}{2} Y^{\prime}_{R}$, where $T^{\prime}_{3R}=\frac{1}{2} T_{3R}+\frac{3}{2}Y_{R}$, $Y^{\prime}_{R}=\frac{1}{2} T_{3R}-\frac{1}{2} Y_{R}$. The superpotential corresponding to the interactions of the SM and new exotic superfields in ALRSM is given by \cite{Hewett:1988xc}
 \begin{eqnarray}
 \label{eq:Wcase1}
 && W= \lambda_1\left( u u^{c} N^{c}_E - d u^{c} E^{c} - u h^{c} e + d h^{c} \nu_e \right)+ \nonumber\\
 &&  \lambda_2 \left( u d^{c} E - d d^{c} \nu_{E}\right)+\lambda_3 \left( h u^c e^c - h h^c n\right) + \nonumber\\
 && \lambda_4 h d^c N^{c}_L +\lambda_5 \left ( ee^c \nu_E + E E^c n - E e^{c} \nu_e- \nu_E N^{c}_E n\right) +\nonumber\\
 && \lambda_6 \left( \nu_e N^{c}_L N^{c}_E - e E^c N^{c}_L\right).
 \end{eqnarray}
 The assignments of $R$-parity, baryon number ($B$) and lepton number ($L$) for the exotic fermions can be readily obtained from the superpotential given in Eq. (\ref{eq:Wcase1}) . For leptoquark $h$ the assignments are $R=-1, B=\frac{1}{3}, L=1$, while the color-neutral $\nu_{E}, E$ and $n$ have the assignments $R=-1, B=L=0$. Their supersymmetric scalar partners have positive $R$ parity assignments. There are two possible choices for the assignments for $N^{c}$. If $N^{c}$ is assigned $R=-1$ and $B=L=0$ (with $\lambda_{4}=\lambda_{6}=0$ in Eq. (\ref{eq:Wcase1})), then $\nu_{e}$ is exactly massless. However if $N^{c}$ is assigned $R=+1$, $B=0$, $L=-1$, then $\nu_{e}$ can acquire a small mass through the seesaw mechanism, which makes this case particularly intriguing from a leptogenesis point of view. 
 \section{Diphoton signal from $\tilde{n}$ decay}
 \label{sec2.1}
 In this section, we argue that gluon-gluon fusion can give the observed production rate of the $750 \gev$ resonance, $\tilde{n}$ in our model, through a loop of scalar leptoquarks ($\tilde{h}^{(c)}$). Subsequently, $\tilde{n}$ decays into $gg$ and $\gamma\gamma$ final states via loops of $\tilde{h}^{(c)}$ and $\tilde{E}^{(c)}$. Note that, considering only scalar leptoquarks in the decay loop of $\tilde{n}$ yields a diphoton branching ratio suppressed by a factor of $10^{-3}-10^{-4}$, and it is the contribution from $\tilde{E}^{(c)}$ loop which enhances the diphoton branching ratio significantly to give the observed cross section of the diphoton signal. 
 
 \begin{figure}[ht!]
   \hspace{0.02cm}
    \hbox{\hspace{0.03cm}
    \hbox{\includegraphics[scale=0.45]{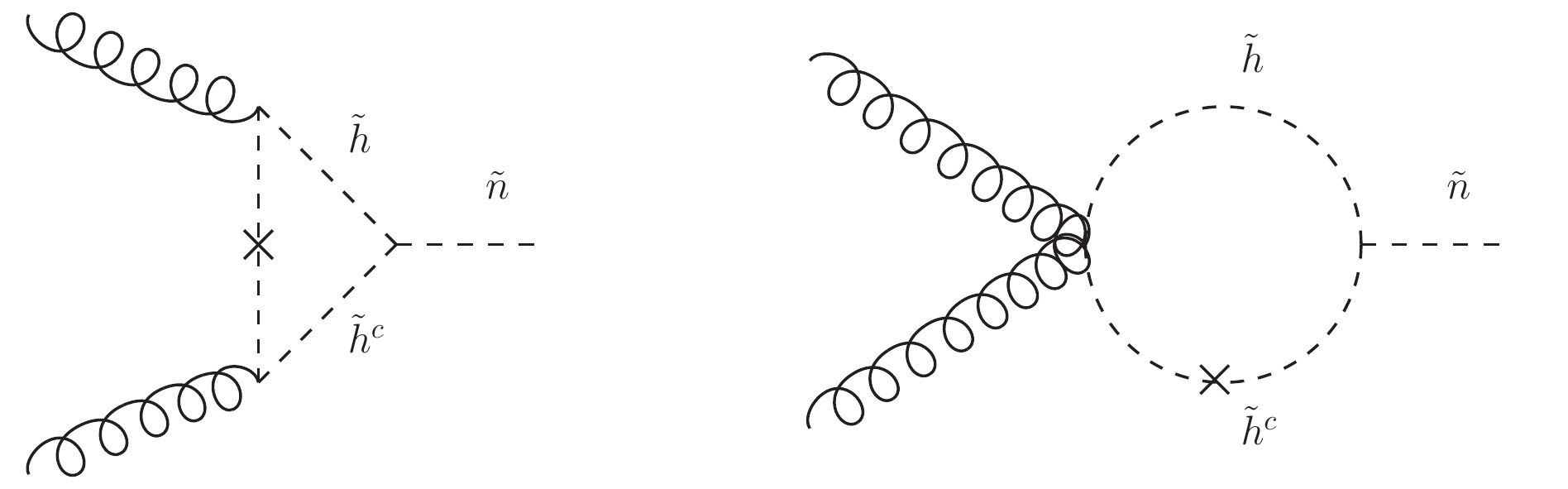}}
    }
    \includegraphics[scale=0.65]{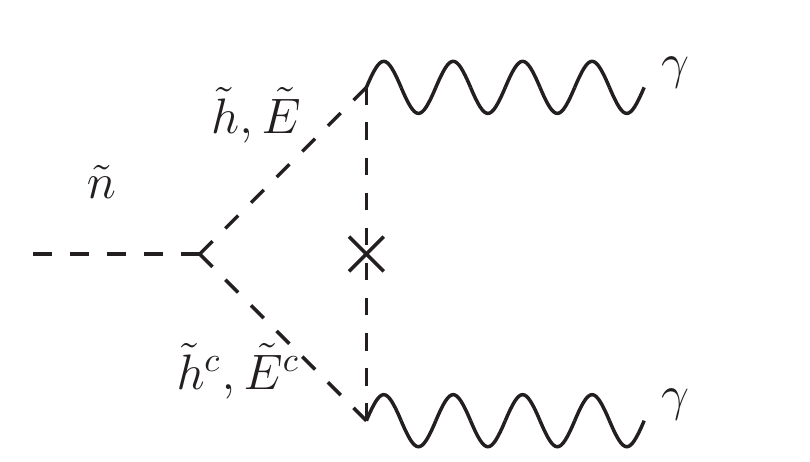}
     \caption{ Production of the scalar resonance $\tilde{n}$ in gluon fusion via scalar leptoquark loops and the subsequent decay into $\gamma\gamma$ final state via loops of scalar letoquark and slepton.}
     \label{fig1}
    \end{figure}
    
\begin{figure}[h]
   \hspace{0.02cm}
    \hbox{\hspace{0.03cm}
    \hbox{\includegraphics[scale=0.40]{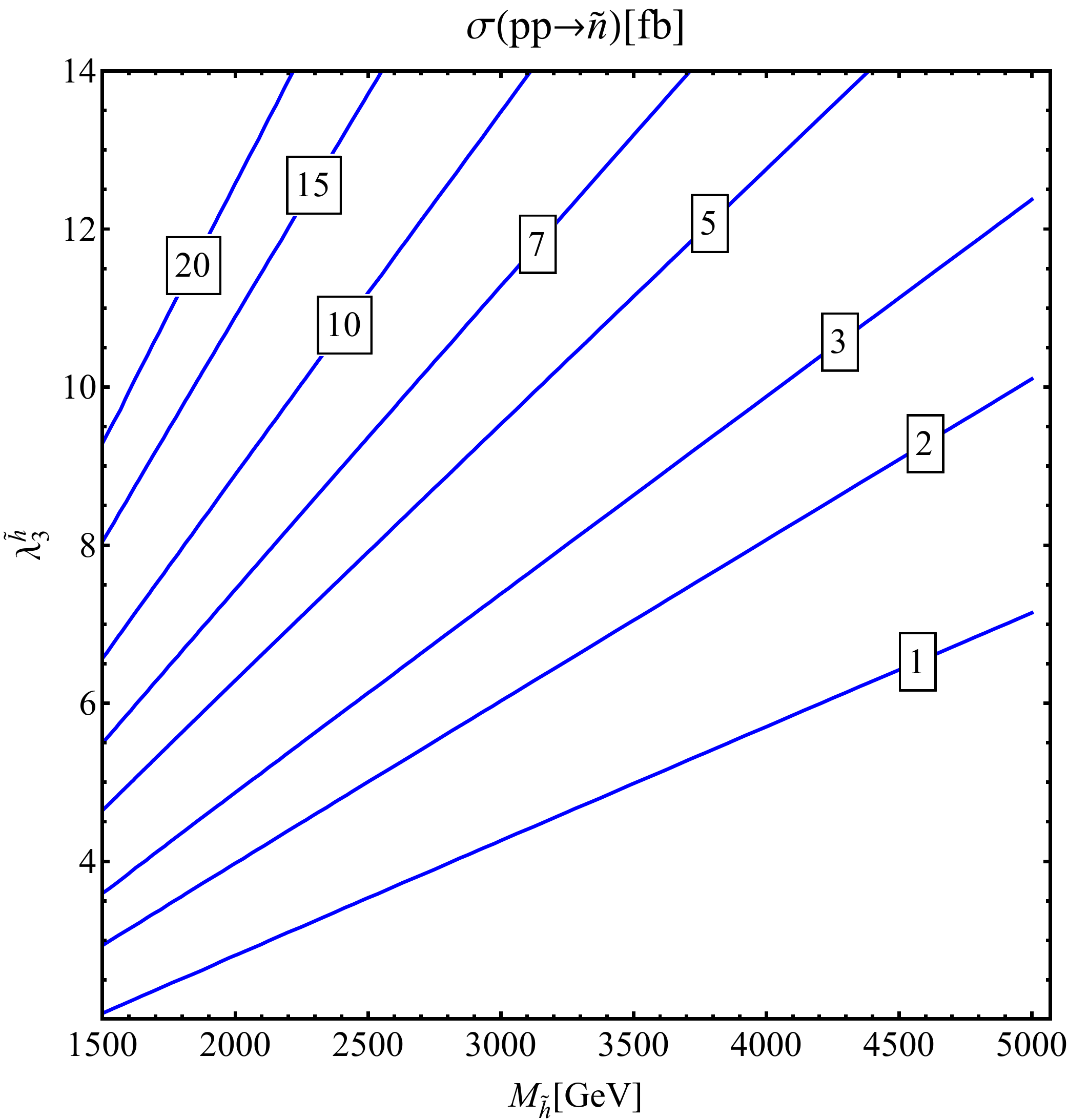}}}
    
    \caption{ $pp\rightarrow \tilde{n}$ production cross section in gluon fusion at $\sqrt{s}=13 \tev$ as function of scalar leptoquark mass and  $\lambda_3^{\tilde{h}}$. }
     \label{fig2}
    \end{figure}
 The relavant terms in the superpotential are given by
  \begin{eqnarray}
 \label{eq:W1}
  W_{1}= -\lambda_3  h h^c n +\lambda_5  E E^c n,
 \end{eqnarray}
 where we have dropped the generation indices for simplicity. The relevant diagrams for $\tilde{n}$ are shown in Fig. \ref{fig1}. The production cross section can be conveniently parametrized in terms of the corresponding production cross section of the SM Higgs $H$ with its mass replaced by the $\tilde{n}$ mass $M_{H}=M_{\tilde{n}}$ \cite{Bauer:2015boy}. This eliminates the factors due to higher order QCD corrections to give 
  \begin{eqnarray}
 \label{2.1}
 \frac{\sigma(pp\rightarrow \tilde{n})}{\sigma(pp\rightarrow H)}= \left(\frac{\lambda_3^{\tilde{h}\tilde{n}} \cos \theta_{\tilde{h}}\sin \theta_{\tilde{h}} v} {8M_{\tilde{h}}} \right)^{2}
 \left| \frac{A_{0}(x_{\tilde{h}})}{A_{1/2}(x_{t})} \right| ^{2},
 \end{eqnarray} 
 where the dimensionful coupling corresponding to the $\lambda_{3}$ trilinear scalar term in Eq. (\ref{eq:W1}) is parametrized as $\lambda^{s}_{3}=\lambda_3^{\tilde{h}\tilde{n}} M_{\tilde{n}}$, $\theta$ is the left-right mixing angle of the scalar leptoquark sector corresponding to $\tilde{h}-\tilde{h}^{c}$, $v$ is the vacuum expectation value $\langle H \rangle=v$, $x_{\tilde{h}}=m_{\tilde{n}}^{2}/4M_{\tilde{h}}^{2}$ and $x_{t}=m_{\tilde{n}}^{2}/4M_{t}^{2}$ where $M_{t}$ is the top mass. The loop functions are given by
  \begin{eqnarray}
 \label{2.1}
A_{0}(x)&=&\frac{3 (f(x)-x)}{x^2},\nonumber\\
A_{1/2}(x)&=&\frac{3}{2x^{2}}\left[ x+(x-1)f(x)\right],
 \end{eqnarray} 
 with $f(x)$ given by
 \begin{eqnarray}{\label{2.2}}
f(x)=\left\{ \begin{matrix}\arcsin^{2} (\sqrt{x}) & & x\leq 1\\ -\frac{1}{4}\left[  \ln \left( \frac{1+\sqrt{1-x}}{1-\sqrt{1-x}} \right)-i\pi\right]^{2}&& x\geq 1.\end{matrix} \right. 
\end{eqnarray}
$\sigma(pp\rightarrow H)$ at $\sqrt{s}=13 \tev$ can be obtained by boosting the $\sqrt{s}=8 \tev$ cross section $\sigma=0.157 \rm{pb}$ (for $M_{H}=750 \gev$) by a factor 4.7 corresponding to increased gluon luminosity \cite{Heinemeyer:2013tqa}. The $pp\rightarrow \tilde{n}$ production cross section in gluon fusion as function of scalar leptoquark mass and  $\lambda_3^{\tilde{h}}=\lambda_3^{\tilde{h}\tilde{n}} \left( M_{\tilde{n}}/M_{\tilde{h}} \right)$ is shown in Fig. \ref{fig2}. Note that, we take the maximum value of $\lambda_3^{\tilde{h}}$ as $14$ corresponding to the rough upper limit from perturbativity \cite{Allanach:2015blv, Cho:2006sm} and $\theta_{\tilde{h}}=\pi/4$ corresponding to maximal mixing between left and right handed scalar leptoquarks.
 
 Now, for $2M_{\tilde{h}(\tilde{E})}>M_{\tilde{n}}$, $\tilde{n}$ can not decay to two on shell $\tilde{h}(\tilde{E})$, giving appreciable branching ratios for $\gamma\gamma$ and $gg$ final states. The partial widths for $\gamma\gamma$ final state are given by
  \begin{eqnarray}{\label{2.3}}
\Gamma^{X}_{\gamma\gamma}=\frac{\alpha^{2}M_{\tilde{n}}^{3}}{256\pi^{3}} \frac{\left|\lambda_{y}^{\tilde{X}\tilde{n}}\cos \theta_{X}\sin \theta_{X}M_{\tilde{n}}\right|^{2}}{M_{X}^{4}}\left| A_{0}(x_{X})\right| ^{2},
\end{eqnarray}
where $X(y)$ can be $\tilde{h}(3)$  and $\tilde{E}(5)$, $A_{0}$ corresponds to the loop function defined in Eq. (\ref{2.1}) and $x_{X}=m_{\tilde{n}}^{2}/4M_{X}^{2}$. The corresponding decay width for $gg$ final state can be obtained by
  \begin{eqnarray}{\label{2.4}}
\Gamma_{gg}=\Gamma^{\tilde{h}}_{\gamma\gamma}\frac{2K_{gg}\alpha_{s}^{2}}{9Q_{\tilde{h}}^{4} \alpha^{2}},
\end{eqnarray}
where $K_{gg}\sim 2$ arises from higher order QCD corrections, $\alpha_{s}(M_{\tilde{n}})\approx 0.092$,$Q_{\tilde{h}}=-1/3$. Considering $\tilde{h}$ as the only field running in the decay loop yields a branching fraction $\sim 10^{-3}-10^{-4}$ for $\gamma\gamma$ final state. Thus, the contribution coming from $\tilde{E}$ running in the final decay loop plays an essential role in controlling the branching ratio to $\gamma\gamma$ final state. The branching fraction as a function of slepton mass $M_{\tilde{E}}$ and scalar leptoquark mass $M_{\tilde{h}}$ is shown in Fig. \ref{fig4}.
\begin{figure}[ht!]
   \hspace{0.02cm}
    \hbox{\hspace{0.03cm}
    \hbox{\includegraphics[scale=0.40]{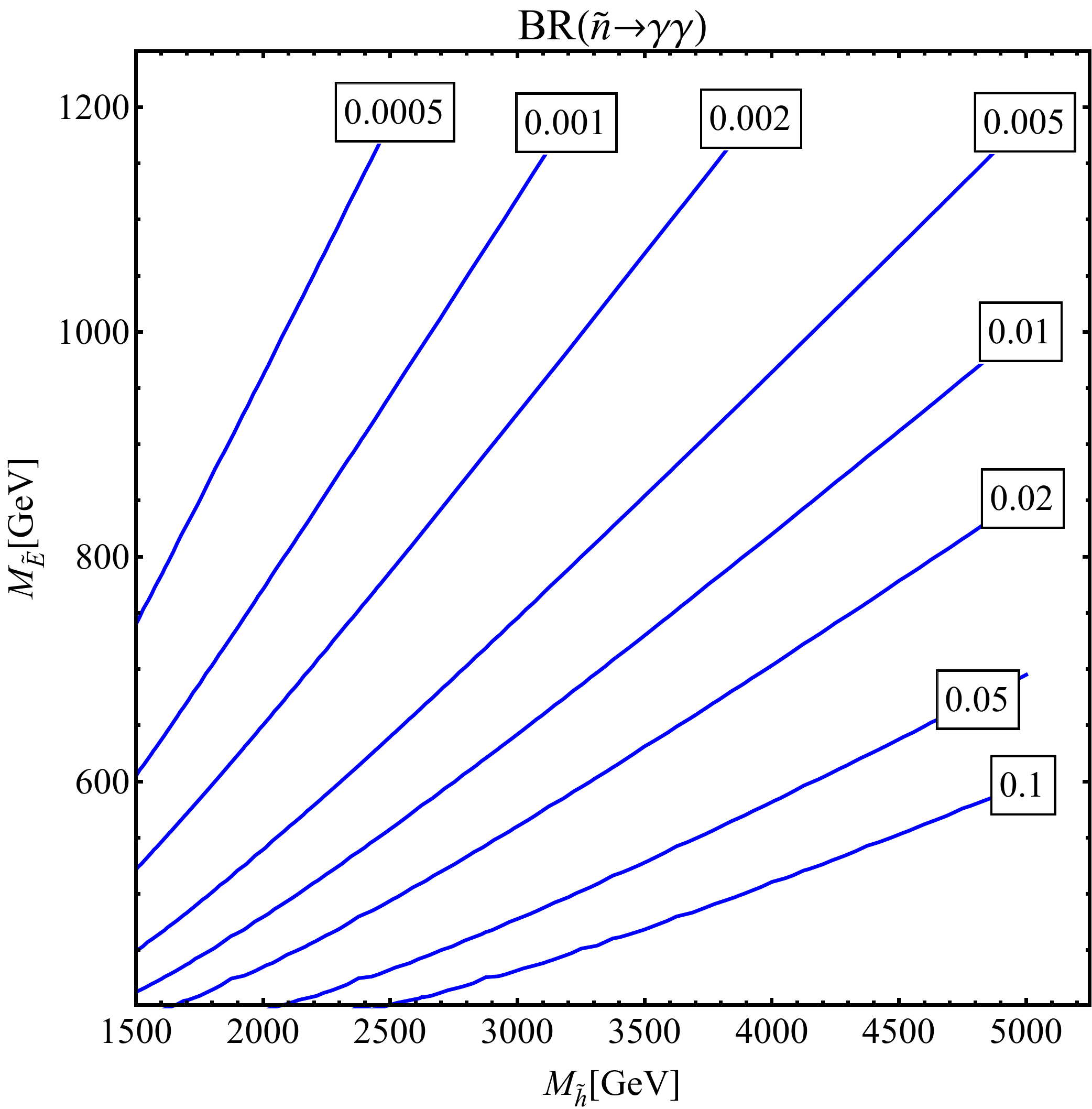}}}
     \caption{ The branching fraction as a function of slepton mass $M_{\tilde{E}}$ and scalar leptoquark mass $M_{\tilde{h}}$ with $\lambda_3^{\tilde{h}}=\lambda_5^{\tilde{E}}=14$. }
     \label{fig4}
    \end{figure}
In Fig. \ref{fig5}, the production cross section times branching ratio $\sigma(pp\rightarrow \tilde{n})\times \rm{BR}(\tilde{n}\rightarrow \gamma\gamma)$ is presented for $\lambda_{3}^{\tilde{h}}=\lambda_{3}^{\tilde{E}}=14$ corresponding to the rough upper limits allowed by perturbativity for the lowest masses of scalar leptoquark ($\tilde{h}$) and slepton ($\tilde{E}$) respectively (${M_{\tilde{h}}}^{\rm{min}}\sim 2 M_{\tilde{n}}$ and ${M_{\tilde{E}}}^{\rm{min}}\sim M_{\tilde{n}}/2$)\footnote{Note that, here we have used the parametrizations $\lambda_3^{\tilde{h}}=\lambda_3^{\tilde{h}\tilde{n}} \left( M_{\tilde{n}}/M_{\tilde{h}} \right)$ and $\lambda_5^{\tilde{E}}=\lambda_5^{\tilde{E}\tilde{n}} \left( M_{\tilde{n}}/M_{\tilde{E}} \right).$} \cite{Allanach:2015blv, Cho:2006sm}, $\theta_{\tilde{h}}=\theta_{\tilde{E}}=\pi/4$ corresponding to maximal mixing between the left and right handed scalar leptoquarks and sleptons. The red band corresponds to the observed value of $\sigma(pp\rightarrow \tilde{n})\times \rm{BR}(\tilde{n}\rightarrow \gamma\gamma)=2-8$ fb, corresponding to 95\% CL upper limit on the allowed cross section at $13 \tev$, consistent with cross section exclusion at $95\%$ CL by the absence of a signal in the CMS run 1 data. We find that slepton mass $M_{\tilde{E}}\lsim 400 \gev$ is favored by the fit, while scalar leptoquark mass $M_{\tilde{h}}\lsim 2500 \gev$ is preferred by the diphoton excess. 

\begin{figure}[ht!]
   \hspace{0.02cm}
    \hbox{\hspace{0.03cm}
    \hbox{\includegraphics[scale=0.50]{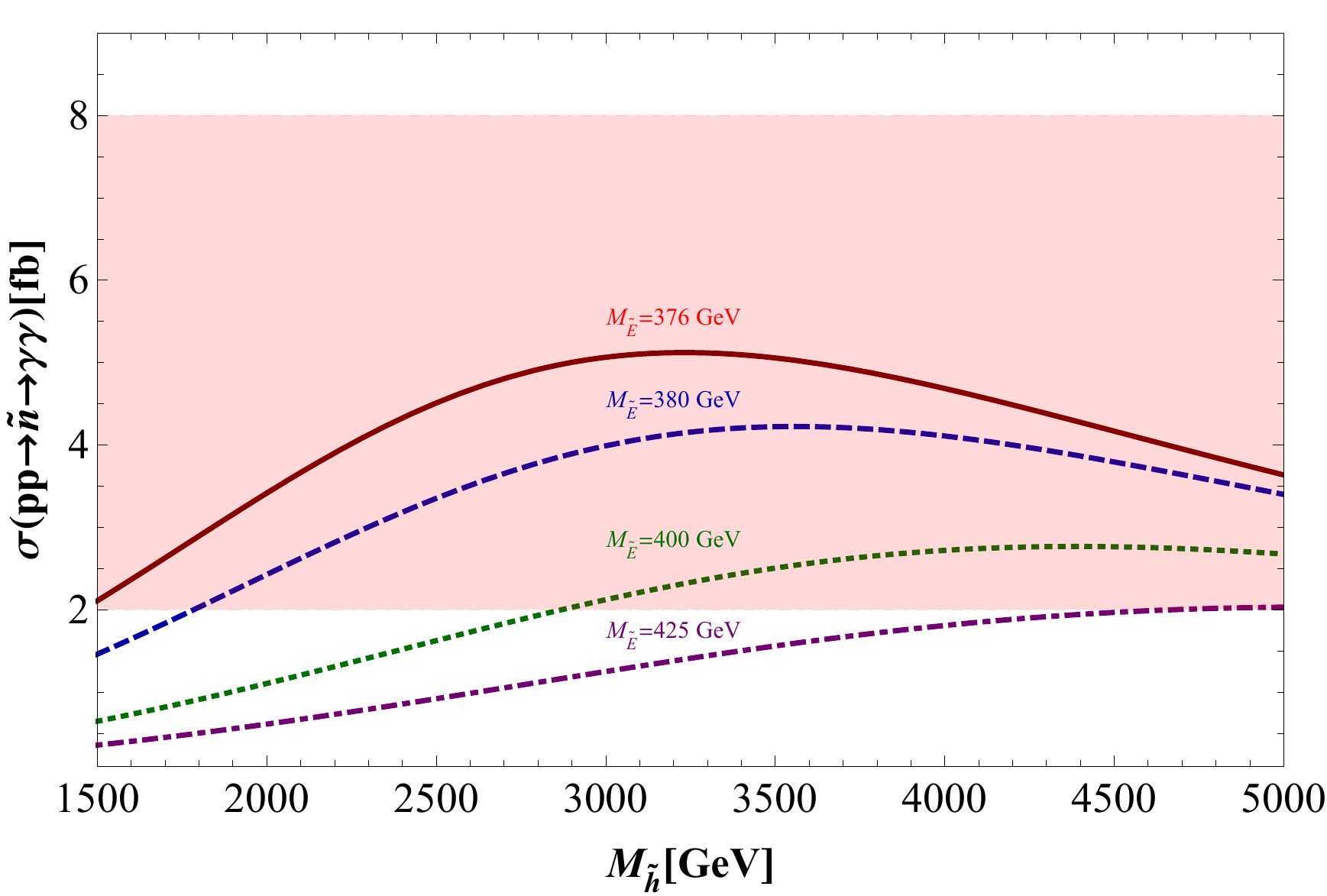}}}
     \caption{ The production cross section times branching ratio $\sigma(pp\rightarrow \tilde{n})\times \rm{BR}(\tilde{n}\rightarrow \gamma\gamma)$ as a function of scalar leptoquark mass $M_{\tilde{h}}$ (for three different values of slepton mass $M_{\tilde{E}}$) with $\lambda_{3}^{\tilde{h}}=\lambda_{3}^{\tilde{E}}=14$, $\theta_{\tilde{h}}=\theta_{\tilde{E}}=\pi/4$. The red band corresponds to the observed value of $\sigma(pp\rightarrow \tilde{n})\times \rm{BR}(\tilde{n}\rightarrow \gamma\gamma)=2-8$ fb, corresponding to $95\%$ CL upper limit on the allowed cross section at $13 \tev$. }
     \label{fig5}
    \end{figure}
Note that, for different generations of $\tilde{n}$ with mass difference $\cO(10)\gev$ one can address the wider peak hinted by ATLAS, given that the present statistics can not resolve these different masses. Thus, if in future if such multiple peak structure is confirmed then it will be interesting to explore this possibility.
\section{Flavor anomalies, other LHC excesses and leptogenesis in ALRSM}
\label{sec3}
In this section, we briefly summarize how ALRSM can explain the enhanced $\bar{B}\rightarrow D^{(\ast)}\tau \bar{\nu}$ rates, the lepton non-universality of the ratio $\cR_{K}=Br(B^{+}\rightarrow K^{+}\mu^{+}\mu^{-})/Br(B^{+}\rightarrow K^{+} e^{+} e^{-})$, the $eejj$ and $e\slashed{p}_{T}jj$ excesses reported by CMS in run 1 of LHC and the possibility of high scale leptogenesis in presence of a $\tev$ scale heavy $W$ boson. As we will discuss below, the scalar leptoquark plays the central role in addressing these anomalies.
\begin{figure}[h]
   \hspace{0.02cm}
    \hbox{\hspace{0.03cm}
    \hbox{\includegraphics[scale=0.25]{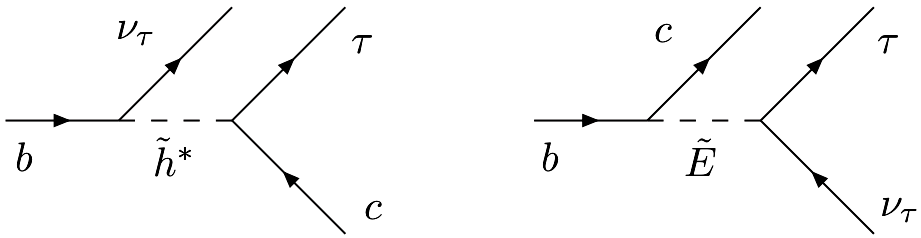}}
    }
    
    \caption{ Feynman diagrams inducing the $\bar{B}\rightarrow D^{(\ast)} \tau \bar{\nu}$ decays in ALRSM by the exchange of scalar leptoquark ($\tilde{h}^{\ast}$) and $\tilde{E}$.
     }
     \label{fig6}
    \end{figure}

From the superpotential given in Eq. (\ref{eq:Wcase1}) it follows that in ALRSM the $\bar{B}\rightarrow D^{(\ast)}\tau \bar{\nu}$ decay rates can be enhanced by the two possible diagrams shown in Fig. \ref{fig6}. The corresponding terms in the effective Lagrangian is given by
\bea
\label{3.1}
\cL_{\rm{eff}}&=&-\sum_{j,k=1}^{3} V_{2k}\left[ \frac{ \l^{5}_{33j} \l^{2\ast}_{3kj}}{M_{\tilde{E}^j}^{2}} \bar{c}_{L}b_{R} \; \bar{\tau}_{R}\nu_{L} \right. \nonumber\\
&&\left. + \frac{ \l^{1}_{33j} \l^{1\ast}_{3kj}}{M_{\tilde{h}^{j \ast}}^{2}} \bar{c}_{L}(\tau^{c})_{R} \; (\bar{\nu}^{c})_{R} b_{L} \right],
\eea
where the superpotential coupling indices are moved to superscripts and the generation indices are shown explicitly as subscripts. $V$ is the CKM matrix. 
The second term of Eq. (\ref{3.1}) can be Fiertz transformed to obtain
\be
\label{3.2}
\bar{c}_{L}(\tau^{c})_{R} \; (\bar{\nu}^{c})_{R} b_{L} = \frac{1}{2} \bar{c}_{L}\gamma^{\mu} b_{L} \; \bar{\tau}_{L} \gamma_{\mu} {\nu}_{L}.
\ee
In the notation of Ref. \cite{Hati:2015awg} the relevant Wilson coefficients are given by
\bea
\label{3.3}
C^{\tau}_{S_L} &=& \frac{1}{2\sqrt{2} G_{F} V_{cb}} \sum_{j,k=1}^{3} V_{2k} \frac{ \l^{5}_{33j} \l^{2\ast}_{3kj}}{m_{\tilde{E}^j}^{2}} \; ,\nonumber\\
C^{\tau}_{V_L} &=& \frac{1}{2\sqrt{2} G_{F} V_{cb}} \sum_{j,k=1}^{3} V_{2k} \frac{ \l^{1}_{33j} \l^{1\ast}_{3kj}}{2\,m_{\tilde{h}^{j \ast}}^{2}},
\eea
where the neutrinos are assumed to be of tau flavor. $C^{\tau}_{S_L}$ contribution alone cannot explain both $\cR(D)$ and $\cR(D^{\ast})$ data simultaneously, however, for $\left| C^{\tau}_{V_{L}} \right| > 0.08$ both $\cR(D)$ and $\cR(D^{\ast})$ data can be simultaneously explained. In a recent study, we have shown that the leptonic decays $D_{s}^{+} \rightarrow \tau^{+} \bar{\nu}$, $B^{+}\rightarrow \tau^{+} \bar{\nu}$, $D^{+}\rightarrow \tau^{+} \bar{\nu}$ and $D^{0}$-$\bar{D}^{0}$ mixing can be used to constrain the couplings involved in the $\bar{B}\rightarrow D^{(\ast)} \tau \bar{\nu}$ decays in ALRSM, and we found that ALRSM can explain the current data on $\cR(D^{(\ast)})$ quite well while satisfying the constraints from the rare $B$, $D$ decays $D^{0}$-$\bar{D}^{0}$ mixing \cite{Hati:2015awg}.
\begin{widetext}
\onecolumngrid
\begin{figure}[ht!]
   \hspace{0.02cm}
    \hbox{\hspace{0.03cm}
    \hbox{\includegraphics[scale=0.7]{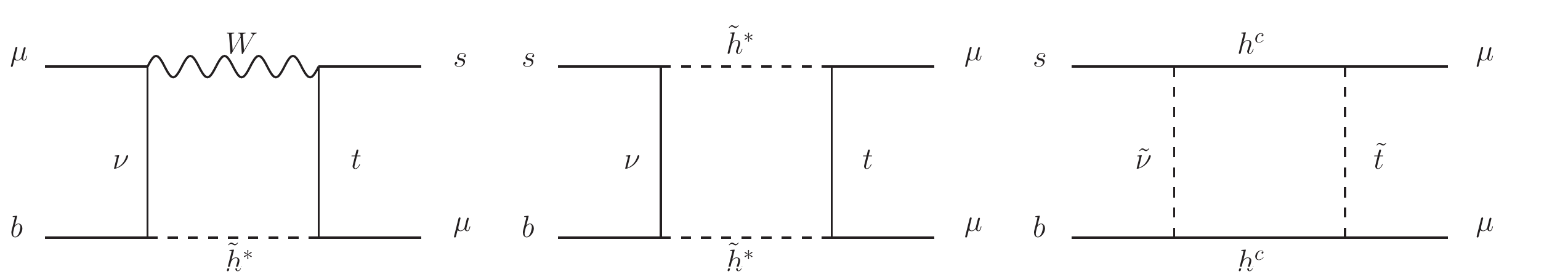}}
    }
     \caption{ Box diagrams inducing $B^{+}\rightarrow K^{+}\mu^{+}\mu^{-}$ decay in ALRSM.
     }
     \label{fig7}
    \end{figure}
        \end{widetext}
    \twocolumngrid

Interestingly, the lepton non-universality of the ratio $\cR_{K}=Br(B^{+}\rightarrow K^{+}\mu^{+}\mu^{-})/Br(B^{+}\rightarrow K^{+} e^{+} e^{-})$ can also be addressed at one loop level where the scalar leptoquarks can naturally induce flavor universality violation via the terms $\lambda_{1}(- u h^{c} e + d h^{c} \nu_e)$ in the superpotential given in Eq. (\ref{eq:Wcase1}). The relevant diagrams are shown in Fig. \ref{fig7}. The upper two vertices of the $W-\tilde{h}^{c}$ box diagram are SM vertices and the amplitude of this diagram is regulated by the coupling of the scalar leptoquark with top and muon. Note that, the $W-\tilde{h}^{c}$ box diagram gives a positive contribution to the Wilson coefficient inducing $b\rightarrow s\mu^{+}\mu^{-}$ given by \cite{Bauer:2015knc}
\bea
\label{3.4}
\left. C_{LL}^{\mu(\tilde{h}^{c})}\right|_{W-\tilde{h}^{c}}=\frac{M_{t}^{2}}{8\pi\alpha M_{\tilde{h}^{c}}^{2}}\left| \lambda^{1}_{23j} \right| ^{2},
\eea
while the other two diagrams give negative contributions proportional to $(\l_{32i} \l_{32j}^{\dagger})(\l_{23i} \l_{23j}^{\dagger})$. A good fit to the current data of $\cR_{K}=Br(B^{+}\rightarrow K^{+}\mu^{+}\mu^{-})/Br(B^{+}\rightarrow K^{+} e^{+} e^{-})$ requires $-1.5 <C_{LL}^{\mu(\tilde{h}^{c})}<-0.7$ \cite{Bauer:2015knc}. Thus the contributions from $\tilde{h}^{c}-\tilde{h}^{c}$ and $\tilde{\nu}-\tilde{t}$ box diagrams are essential to explain the non-universality of $\cR_{K}$.

 ALRSM can also explain both $eejj$ and $e\slashed p_T jj$ signals, reported by CMS during run 1 of LHC, through (i) resonant production of the slepton $\tilde E$, which then subsequently decays into a charged lepton and a neutrino, followed by a structure similar to neutrinoless double beta decay, producing an excess of events in both  $eejj$ and $e\slashed p_Tjj$ channels \cite{Dhuria:2015hta} (ii) pair production of scalar leptoquarks ${\tilde h}$. However, given that a good fit to the diphoton signal data requires a low mass $\tilde{E}$ ($M_{\tilde{E}}\lsim 400\gev$), the latter scenario is clearly more favorable.
 
    \begin{figure}[ht]
   \hspace{0.02cm}
    \hbox{\hspace{0.03cm}
    \hbox{\includegraphics[scale=0.35]{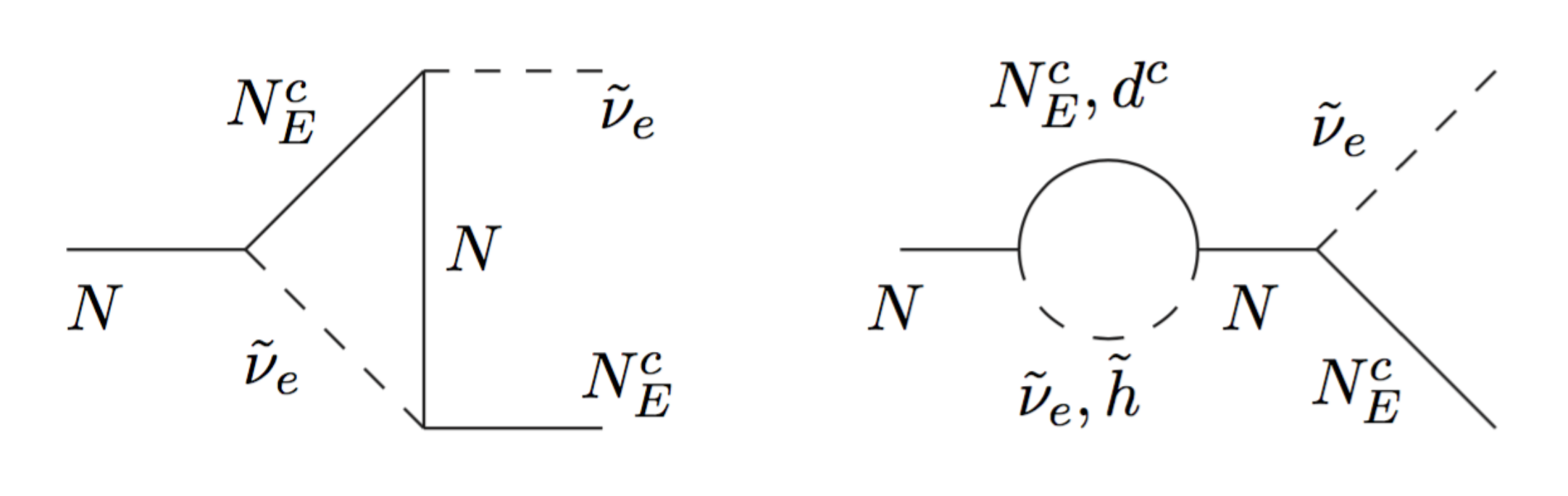}}
    }
     \caption{ One loop diagrams for $N$ decay interfering with tree level diagram inducing $CP$ violation.
     }
     \label{fig8}
    \end{figure}
 Finally, ALRSM can also accommodate the attractive possibility high scale leptogenesis, which can address the observed baryon asymmetry of the universe. In the conventional LRSM, the observation of a $2\tev$ $W_{R}$ (through the confirmation of the $eejj$ signal in $W_{R}$ search reported by the CMS) rules out the possibility of high scale as well as $\tev$ scale resonant leptogenesis due to fast gauge mediated $B-L$ violating interactions involving $W_{R}$ \cite{Ma:1998sq,Frere:2008ct,Deppisch:2015yqa,Deppisch:2013jxa,Dev:2014iva,Dhuria:2015wwa,Dhuria:2015cfa,Dev:2015vra}. ALRSM provides a way around this problem, where  high scale leptogenesis can be obtained via the decay of the complete gauge singlet heavy Majorana neutrino $N^c$. From the interaction terms $\lambda_{4}$ and $\lambda_{6}$ in Eq. (\ref{eq:Wcase1}), it follows that $N^c_{k}$ can decay into the final states $\nu_{e_{i}} {\tilde N}^{c}_{E_{j}}, {\tilde \nu}_{e_{i}} N^{c}_{E_{j}}, e_{i}{\tilde E}^{c}_{j}, {\tilde e}_{i}, E^{c}_{j}$ and $d_{i} {\tilde h}_{j}, {\tilde d^{c}}_{i} {\tilde h}_{j}$ with total $B-L=-1$ and to their conjugate states with total $B-L=+1$. The corresponding one loop diagrams interfering with tree level decay diagrams to induce $CP$ violation are shown in Fig. \ref{fig8}. For $M_{N}\sim 10^{15} \gev$, $\l^{4,6}_{ijk}\sim 10^{-3}$ gives the observed baryon on entropy ratio $n_{B}/s \sim 10^{-10}$ for a maximally $CP$ violating scenario \cite{Dhuria:2015hta}. Thus, from leptogenesis point of view also ALRSM is a very interesting framework.

\section{Conclusions}
\label{sec4}
We have studied the possibility of explaining the diphoton signal in $E_6$ motivated Alternative Left-Right Symmetric Model, capable of explaining the $B$ decay anomalies , the $eejj$ and $e\slashed{p}_{T}jj$ excesses reported by CMS in run 1 of LHC and high scale leptogenesis. We found that gluon-gluon fusion can give the observed production rate of the $750 \gev$ resonance, $\tilde{n}$, through a loop of scalar leptoquarks ($\tilde{h}^{(c)}$) with mass below a few \tev range. $\tilde{n}$ can subsequently decay into $gg$ and $\gamma\gamma$ final states via loops of $\tilde{h}^{(c)}$ and $\tilde{E}^{(c)}$. Considering only scalar leptoquarks in the decay loop of $\tilde{n}$ yields a diphoton branching ratio suppressed by a factor of $10^{-3}-10^{-4}$, however, the contribution from $\tilde{E}^{(c)}$ loop enhances the diphoton branching ratio significantly can explain the observed cross section of the diphoton signal. We have also discussed the possibility of explaining the enhanced $\bar{B}\rightarrow D^{(\ast)}\tau \bar{\nu}$ decay rates, the lepton non-universality of the ratio $\cR_{K}=Br(B^{+}\rightarrow K^{+}\mu^{+}\mu^{-})/Br(B^{+}\rightarrow K^{+} e^{+} e^{-})$, the $eejj$ and $e\slashed{p}_{T}jj$ excesses and high scale leptogenesis in ALRSM. If these excess signals are confirmed in future at the LHC and future $B$-physics experiments then ALRSM will be one of the interesting candidates for new physics beyond the Standard Model.


\bigskip\bigskip
\section*{Acknowledgment}
The author would like to thank Utpal Sarkar, Namit Mahajan, Raghavan Rangarajan, Mansi Dhuria and Girish Kumar for many helpful discussions.  
 
\end{document}